\documentclass[fleqn,usenatbib]{mnras}

\usepackage{newtxtext,newtxmath}

\usepackage[T1]{fontenc}
\usepackage[dvipsnames, usenames]{xcolor}

\DeclareRobustCommand{\VAN}[3]{#2}
\let\VANthebibliography\thebibliography
\def\thebibliography{\DeclareRobustCommand{\VAN}[3]{##3}\VANthebibliography}

\usepackage{graphicx}	
\usepackage{amsmath}	

\title[binary mergers]{Circumbinary disc self-gravity governing supermassive black hole binary mergers}

\author[A. Franchini, A. Sesana and M. Dotti]{
Alessia Franchini,$^{1,2}$\thanks{E-mail: alessia.franchini@unimib.it}
Alberto Sesana,$^{1}$
Massimo Dotti,$^{1,2}$
\\
$^{1}$ Dipartimento di Fisica ``G. Occhialini", Universit\'a degli Studi di Milano-Bicocca, Piazza della Scienza 3, 20126 Milano, Italy \\
$^{2}$ INFN, Sezione di Milano-Bicocca, Piazza della Scienza 3, 20126 Milano, Italy
}

\date{Accepted XXX. Received YYY; in original form ZZZ}

\pubyear{2020}

\begin{document}
\label{firstpage}
\pagerange{\pageref{firstpage}--\pageref{lastpage}}
\maketitle

\begin{abstract}

Understanding the interaction of massive black hole binaries with their gaseous environment is crucial since at sub-parsec scales the binary is too wide for gravitational wave emission to take over and to drive the two black holes to merge. We here investigate the interaction between a massive black hole binary and a self-gravitating circumbinary disc using 3D smoothed particle hydrodynamics simulations.  We find that, when the disc self-gravity regulates the angular momentum transport, the binary semi-major axis decreases regardless the choice of disc masses and temperatures, within the range we explored. 
In particular, we find that the disc initial temperature (hence the disc aspect ratio) has little effect on the evolution of the binary since discs with the same mass self-regulate towards  the same temperature. Initially warmer discs cause the binary to shrink on a slightly shorter timescale until the disc has reached the self-regulated equilibrium temperature.
More massive discs drive the binary semi-major axis to decrease at a faster pace compared to less massive discs and result in faster binary eccentricity growth even after the initial-condition-dependent transient evolution.
Finally we investigate the effect that the initial cavity size has on the binary-disc interaction and we find that, in the self-gravitating regime, an initially smaller cavity leads to a much faster binary shrinking, as expected. 
Our results are especially important for very massive black hole binaries such as those in the PTA band, for which gas self gravity cannot be neglected.

\end{abstract}




\section{Introduction}
\label{sec:intro}

Understanding the cosmic evolution of massive black holes (MBHs, with masses between millions and billions solar masses) is one of the key challenges of present day astronomy. 
Observational evidence suggests that such MBHs inhabit the nuclei of (virtually all) massive galaxies \citep{Kormedy2013}.
When two galaxies merge, the MBHs hosted in their nuclei migrate to the center of the merger remnant primarily due to dynamical friction \citep{Chandrasekhar1943} against the background of stars and gas. At parsec separations (the precise value depending on the mass of the MBHs and the density of stars and gas in the host nucleus) the two black holes start to feel each other gravity binding into a massive black hole binary (MBHB). At this point, dynamical friction becomes inefficient and further evolution of the binary requires a physical mechanism able to extract its energy and angular momentum \citep[see][for a review]{Dotti2012}. The main mechanisms proposed in the literature are three-body scattering of stars intersecting the binary orbit \citep[e.g.,][]{Quinlan1996,sesana2007} or the interaction with a circumbinary gaseous disc \citep{Mayer2007b,escala2005,dotti2007,Cuadra2009}.

Since both galaxies might contain large amounts of gas, it is expected that this will sink to the centre of the newly formed galaxy and form a circumbinary accretion disc \citep{begelman1980,escala2005,Cuadra2009}. 
The presence of such a disc around a massive black hole binary might facilitate its merger and potentially give rise to observational signatures in the form of electromagnetic signals. This was first proposed as a potential way to solve the so-called \lq last parsec\rq \,problem \citep{begelman1980,mm2001} by \cite{ArmitageNatarajan2002} and further investigated in \cite{lodato2009}.

In the standard picture of an accretion disc whose angular momentum transport can be modeled with a phenomenological viscosity, the gaseous disc spreads inwards and outwards owing to the presence of viscosity, which is typically modelled with the viscous stress proportional to the local pressure \citep{ss1973,Pringle1991,artymowicz1994}. 
The physical process that drives the viscosity in the disc is usually assumed to be turbulence, either generated by the magneto-rotational instability (MRI) in highly ionized gaseous discs \citep{balbushawley1991} or by gravitational instabilities (GIs) \citep{paczynski1978,lodato2007sg} in colder or not magnetized discs. In this work we are going to focus on the latter.
A coplanar accretion disc can extend down to a few times the binary separation \citep{artymowicz1994}. The disc orbits resonate with the binary orbit at discrete locations (outer Lindblad resonances), leading to the exchange of angular momentum between the disc and the binary \citep{lynden-bell1972,Lin1986}. The magnitude of the resonant torques depend on the binary potential, i.e. its mass ratio and eccentricity, and are proportional to the disc surface density at the resonance locations \citep{goldreich1979}. Therefore the amount of angular momentum transferred from the binary to the disc at the resonances depends on the disc properties. 

If the binary can transfer a significant amount of angular momentum to small radii inside the disc, the disc can efficiently be truncated and therefore only a small amount of material is able to enter the disc cavity and to accrete onto the binary. Very early numerical simulation works discussed this possibility, finding that it ultimately leads to the binary shrinkage \citep{artymowicz1994,artymowicz1996}. More recent studies concluded that the binary potential might not prevent the gas from leaking inside the cavity \citep{roedig2012,shi2012,dorazio2013,farris2014}.
The fact that time dependent streams of material are able to flow from the disc into the cavity implies that the ratio of outward viscous angular momentum flux from the inner boundary to the inward advected flux is $f\gg 1$ \citep{nixonpringle2020}. 
For discs in which the viscous torque is comparable to the torque produced by the resonances, a non-axisymmetric cavity forms and the binary is fed through streams of material (this corresponds to $f\sim1$). If the disc viscosity is sufficiently high (thick discs), there might be not be enough build up of material at the resonances that can hold the accretion flow from plunging onto the binary ($f\ll1$).

Recently, a number of works employing 2D static or moving-mesh grid numerical simulations with fixed binary orbits have shown that the overall torque exerted onto the binary is positive, meaning that the binary components tend to be drifted apart from each other by the interaction with the gaseous disc \citep{Miranda2017,Moody2019,Duffell2019,Munoz2019,Munoz2020}. 
Using a similar method (i.e. 2D static mesh numerical simulations), \cite{tiede2020} investigated the effect of the disc temperature on this behaviour, finding binary shrinkage for disc aspect ratios below $H/R=0.04$ and expansion for warmer discs.
\cite{heathnixon2020} showed, using 3D Smoothed Particle Hydrodynamics (SPH) simulations, that these conclusions on the binary semi-major axis evolution are extremely sensitive to the choice of the disc and binary parameters. 
These parameters determine whether the resonant torque, which transfers energy and angular momentum to the disc from the binary, can overcome the torque that transfers angular momentum from the material that flows from the circumbinary disc to the binary orbit. 
Nonetheless, they confirm the main conclusion of \cite{tiede2020} that there is some critical disc aspect ratio below which the binary shrinks. However, \cite{heathnixon2020} conclude that the critical aspect ratio for binary expansion, i.e.  $(H/R)_{\rm crit}\approx 0.2$, is much higher than the value $H/R\sim0.04$ reported in \cite{tiede2020}. 

All the above studies did not include the effect of the disc self-gravity, which is likely non negligible for circumbinary discs around massive ($M>10^7 M_{\odot}$) black hole binaries \citep{Cuadra2009,roedig2012}. Understanding the fate of these massive systems is particularly important to forecast the potentially observable population of sub-parsec binaries in the low redshift universe, both via forthcoming time domain surveys \citep[e.g. with the Vera Rubin observatory][]{LSST2009} and ongoing pulsar timing array (PTA) experiments \citep{Reardon2016,Desvignes2016,Verbiest2016,Alam2021}. 
Binaries in the mass range $10^8M_{\odot}-10^{10}M_{\odot}$ are expected to be the loudest gravitational wave (GW) sources in the Universe \citep{Sesana2008}, and their signal might be strong enough to appreciably affect the time of arrival of the pulses from the most stable millisecond pulsars, monitored by PTA experiments within our Galaxy. PTAs, however, are only sensitive to GWs in the nano-Hz frequency range ($\nu\sim 1-100$ nHz) , that are emitted by MBHBs reaching separations $\lesssim 0.01-0.1$ pc (depending on binary mass). It is therefore clear that any dynamical process operating on larger scales leading to the expansion of the orbit will prevent binaries to enter the PTA frequency band, and needs to be carefully considered when modeling the expected overall GW signal from a cosmic population of MBHBs.


As a first step to address this problem, in this paper we study the evolution of equal mass, circular binaries in massive discs. In Section \ref{sec:sg} we first explore the parameter space to understand at which scales the binary evolution is dominated by its interaction with the circumbinary disc and for which parameters we can expect the disc self-gravity to be important. Binaries with separation smaller than a certain threshold will coalescence through GW emission no matter the result of their interaction with the circumbinary disc. We then investigate in Section \ref{sec:hydro} the effect of the disc self-gravity in the picture outlined by \cite{heathnixon2020}, performing SPH simulations in the region of the parameter space where the binary evolution is determined by its interaction with a self-gravitating circumbinary disc. In particular we change the disc temperature, mass and radial extent, exploring the effect that each of this parameters has on the binary evolution.
In all simulations, we find that the effect of a massive circumbinary disc is to shrink the binary semimajor axis, confirming earlier results from \cite{Cuadra2009,roedig2012}. These results and their implications are discussed in Section \ref{sec:discussion}.

\section{Disc self-gravity}
\label{sec:sg}

We start by defining the portion of MBHB mass-separation parameter space where disc self-gravity is relevant and needs to be taken into account in the modeling.

According to the theory of general relativity, the binary semi-major axis decreases as a result of energy and angular momentum carried away by gravitational waves \citep{Peters1964}.
The merger is expected to occur on a timescale 
\begin{equation}
    t_{\rm GW} = \frac{5\,c^5}{256\,G^3}\frac{a^4}{M_1M_2\,M\,F(e_{\rm b})}
    \label{eq:tgw}
\end{equation}
where $a$ is the binary separation, $M_1$ and $M_2$ the primary and secondary black hole mass respectively and $M=M_1 + M_2$ the binary mass. The function of the binary eccentricity $F(e_{\rm b})$ is, to 4th order in $e_{\rm b}$, 
\begin{equation}
    F(e_{\rm b}) = (1-e_{\rm b}^2)^{-7/2}\left(1+\frac{73}{24}e_{\rm b}^2 + \frac{37}{96}e_{\rm b}^4\right)\,.
    \label{eq:fecc}
\end{equation}
We here assume the binary mass ratio to be $q=1$.
This timescale needs to be shorter than the Hubble time, therefore placing a lower limit on the binary semi-major axis $a_{\rm GW}$.

\begin{figure*}
    \includegraphics[width=\columnwidth]{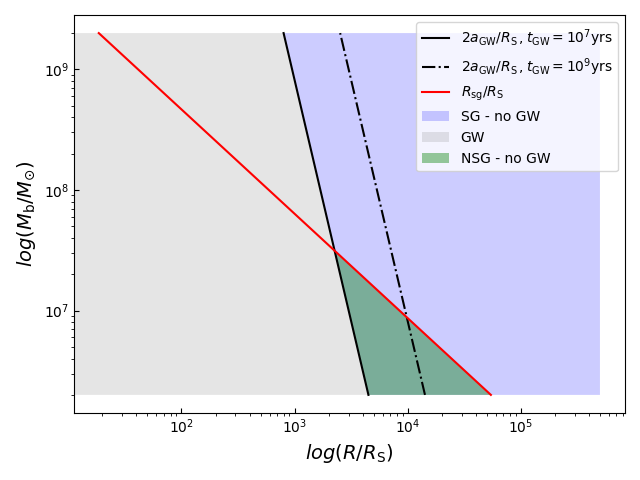}
    \includegraphics[width=\columnwidth]{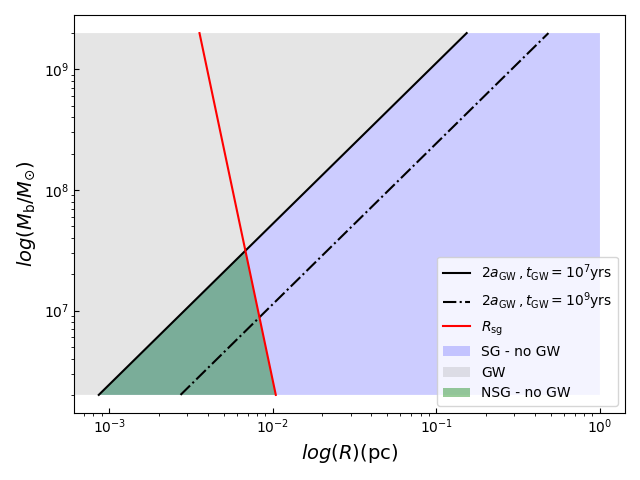}
    \includegraphics[width=\columnwidth]{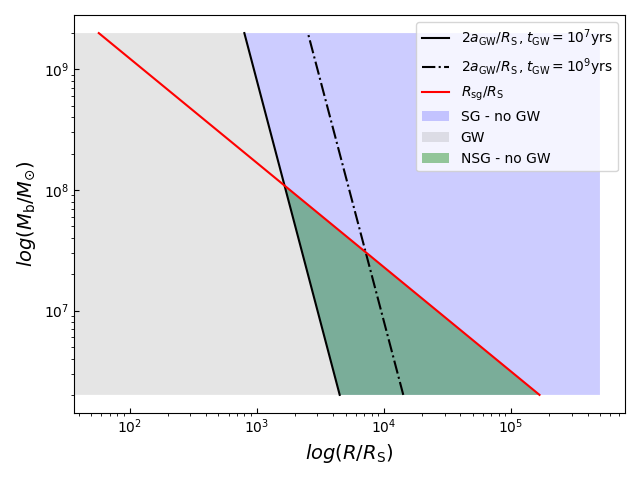}
    \includegraphics[width=\columnwidth]{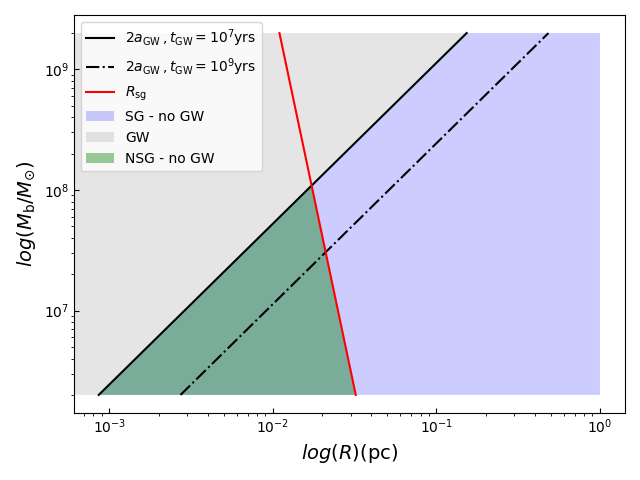}
    \caption{Parameter space for $f_{\rm Edd}=1$ (upper panels) and $f_{\rm Edd}=0.1$ (lower panels). The left panels are in units of $R_{\rm S}$ while the right panels are in pc.}
    \label{fig:paramspace}
\end{figure*}

Wider binaries cannot merge only through gravitational waves emission and therefore their interaction with a circumbinary disc is crucial to bring them to coalescence.
The circumbinary accretion disc  is expected to extend from $\sim 2a$ outwards and we are interested in the regime where $a>a_{\rm GW}$ and the evolution of the binary is not already driven by GW emission.
The solid and dot-dashed line in Figure \ref{fig:paramspace} represent the results obtained with $t_{\rm GW}=10^7$ yrs and $t_{\rm GW}=10^9$ yrs respectively.

The gravitational stability of the disc can be defined, for a Keplerian accretion disc, by the parameter \citep{toomre1964}
\begin{equation}
    Q = \frac{c_{\rm s}\Omega}{\pi G \Sigma}\,.
    \label{eq:toomre}
\end{equation}

The disc self-gravity becomes important at a typical distance $R_{\rm sg}$ where $M_{\rm d}(R_{\rm sg})\sim (H/R)M$ \citep{pringle1981,lodato2007sg}.
Following \citep{perego2009}, we
estimate the self-gravitating radius to be

\begin{equation}
R_{\rm sg} = 1.21\times10^5 \alpha^{28/45}_{0.1} \left(\frac{f_{\rm Edd}}{\eta_{0.1}}\right)^{-22/45}M^{-52/45}_6\,R_{\rm S}   
\label{eq:Rsg}
\end{equation}
where $\alpha$ is the disc vicosity coefficient \citep{ss1973}, $f_{\rm Edd}=\dot{M}/\dot{M}_{\rm Edd}$, $\eta$ is the accretion efficiency, $M_6 = M/10^6M_{\odot}$ and $R_{\rm S}$ is the Schwarzschild radius.

The result is represented by the solid red line in all the panels of Fig. \ref{fig:paramspace}. The upper and lower panels represent Eddington ($f_{\rm Edd}=1.0$) and sub-Eddington accretion ($f_{\rm Edd}=0.1$) respectively.
We have coloured in green the region where the interaction of the binary with a non-self-gravitating accretion disc is important. The blue region identifies where the binary interacts with a self-gravitating accretion disc, and this is the region of interest for this work. The grey area, which extends to every value on the left of the black solid line, represents the region where the gravitational waves emission brings the binary to merge regardless of its interaction with the circumbinary disc. 
Note that we plotted $2a_{\rm GW}$ since this is the expected tidal truncation radius for a coplanar circumbinary disc \citep{artymowicz1994}.

We assumed the binary to have a circular orbit to produce Fig. \ref{fig:paramspace}, i.e. $F(e_{\rm b})\approx 1$ from Eq. \ref{eq:fecc}. However it is worth mentioning that a non-zero binary eccentricity would lead to a slight shift of the solid and dot-dashed black lines towards higher radii. This essentially implies an even narrower region of the parameter space where the disc self-gravity is negligible, at least for massive black hole binaries above $10^7\,M_{\odot}$. 

Note that Eq. \ref{eq:Rsg} assumes that the disc can be described by the \cite{ss1973} model and is therefore geometrically thin.  This assumption begins to break down when $f_{\rm Edd}\gtrsim 1$.
We should therefore in principle consider a disc model with a slim structure inside the photon trapping radius and a Shakura-Sunyaev structure beyond it \citep{Ohsuga2002}. However, photon trapping effects play an important role when $H/R$ is of the order of unity while circumbinary discs surrounding massive black hole binaries are expected to have aspect ratios $\sim 10^{-2}-10^{-3}$ at the scales where the dynamics is not dominated by GWs.

From Fig. \ref{fig:paramspace} we can see that for black hole binaries in the PTA band, i.e. $10^8-10^9\,M_{\odot}$, at separations of the order of $0.01-0.1$ pc, the circumbinary disc is likely to be self-gravitating.
Conversely, binaries with $M<10^7,M_{\odot}$, which are anticipated to be the primary GW sources of the Laser Interferometer Space Antenna (LISA, \cite{AmaroSeoane2017}), are expected to evolve in much lighter discs, and self gravity can safely be neglected in the modeling.

\section{3D Hydrodynamical Simulations of self-gravitating circumbinary discs}
\label{sec:hydro}

\begin{table}
\centering
    \caption{Parameters of the circumbinary disc for each simulation. The first column is the simulation ID. The second, third and fourth columns contain the initial disc inner edge, the aspect ratio at the inner edge and the initial disc mass respectively. The last column contains the initial minimum value of the Q parameter. }
	\begin{tabular}{lcccccll} 
	\hline
    ID  & $R_{\rm in}$ & $H/R(R_{\rm in})$ & $M_{\rm d}/M$ & $Q_{\rm min}$  \\
	\hline
   	sg3b  & $2a$ & 0.1 & 0.1 & 1.48 \\
   	sg4b  & $2a$ & 0.1 & 0.05 & 2.96 \\
   	sg6  & $3a$ & 0.05 & 0.1 & 0.8 \\
   	sg6b  & $2a$ & 0.05 & 0.1 & 0.8 \\
   	sg7b  & $2a$ & 0.2 & 0.1 & 2.96 \\
   	sg8b  & $2a$ & 0.1 & 0.2 & 0.8 \\
	\hline
	\end{tabular}
    \label{tab:sim}
\end{table}

The presence of a non-negligible self-gravity contribution in an accretion disc can either result in the formation of spirals in the disc, that provide a source of angular momentum transport \citep{lodatorice2004,Lodatorice2005}, or in its fragmentation into stars.

Whether the disc outer parts can fragment depends on the choice of the cooling mechanism. For a particle $i$ with specific internal energy $u_{\rm i}$, the cooling is implemented using 
\begin{equation}
\frac{du_{\rm i}}{dt} = -\frac{u_{\rm i}}{t_{\rm cool}} 
\end{equation}
where the details of the cooling function are incorporated into the parameter $t_{\rm cool}$. 
We here use a constant ratio between the dynamical time and the cooling time $\beta_{\rm cool}=\Omega t_{\rm cool}$.
This type of cooling has been used extensively in modelling self-gravitating accretion discs \citep{gammie2001,lodatorice2004,cossins2009,Cuadra2009}.
\cite{gammie2001} found the threshold for fragmentation to be $\beta_{\rm frag}\sim3-4$ with a polytropic index $\gamma=2$, while \cite{Rice2005} found $\beta_{\rm frag}\approx 6$ using $\gamma=5/3$. \cite{cossins2009} found the value of $\beta$ to be in the range $4<\beta_{\rm frag}< 5$ for a disc with surface density profile $\Sigma \propto R^{-3/2}$, slightly different from the employed $\Sigma \propto R^{-1}$ in previous studies. As the polytropic index decreases, discs are more unstable against fragmentation and the critical value of $\beta$ increases \citep{Rice2005}.
We choose $\beta_{\rm cool}=10$ and $\gamma=5/3$ so that the disc is allowed to become sufficiently unstable to form large spirals but not to fragment, since for this work we are not interested in the fragmentation regime. 

\subsection{Numerical setup}
\label{sec:setup}

We use the smoothed particle hydrodynamics (SPH) code {\sc phantom} \citep{Price2010,phantom2017} to model the system composed by an initially equal mass circular binary surrounded by a gaseous self-gravitating accretion disc. 
We consider only coplanar prograde discs in this work. 
Misaligned and retrograde non-self-gravitating circumbinary discs have been investigated using the same code in \citep{nixon2012,nixon2013,lubow2015,franchini2019b}.

In simulation units the initial mass of the binary is $M=M_1+M_2=1$ while its separation is set to $a=1$.
Both binary components are modelled as sink particles with accretion radii $R_{\rm acc}=0.05a$ \citep{bate1995}. Particles inside these radii are accreted onto the respective sink particle ensuring that mass and linear momentum of the system are conserved.
The circumbinary disc initially extends from $R_{\rm rin}=2a$ up to an outer radius $R_{\rm out}=10a$. 

We model the $N=10^6$ particles as a perfect gas using an adiabatic equation of state with $\gamma=5/3$ so that the disc is allowed to be heated up by $PdV$ work and shocks. 
We apply the artificial viscosity only to approaching particles in order to be able to resolve shocks, using the \cite{cullen2010} switch. 
The shock capturing dissipation terms are included in the code according to the approach outlined in \cite{monaghan1997} and consist of a linear term $\alpha_{\rm AV}$, controlled by the switch, and a quadratic term $\beta_{\rm AV}$, which essentially prevents particle interpenetration \citep{lattanzio1986,monaghan1992}.
Since we want the transport of angular momentum induced by gravitational instabilities to dominate, we minimize the numerical dissipation introduced by the numerical viscosity by setting $\alpha_{\rm AV}$ in the range $[0,1]$ and $\beta_{\rm AV}=2.0$ \citep{meru2012}.

We explore a variety of initial disc temperatures, masses and cavity inner edge radii (see Table \ref{tab:sim}) and in all the cases we initialize the disc to be gravitationally stable with $Q\gtrsim 1$ (see last column in Table \ref{tab:sim}).

\subsection{Torques contribution to binary angular momentum evolution}
\label{sec:torques}

\begin{figure*}

    \includegraphics[width=0.33\textwidth]{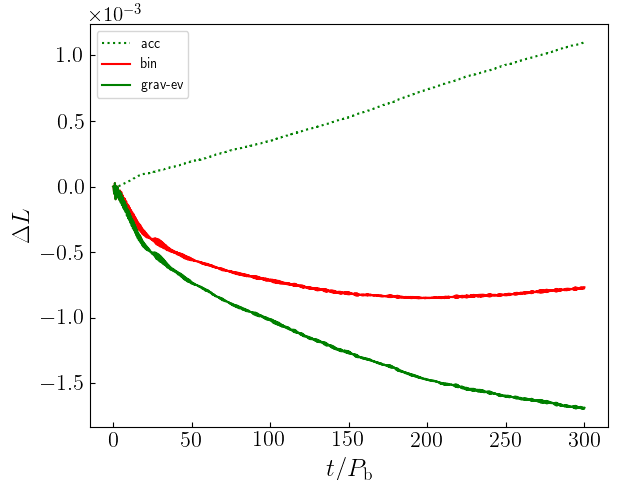}
    \includegraphics[width=0.33\textwidth]{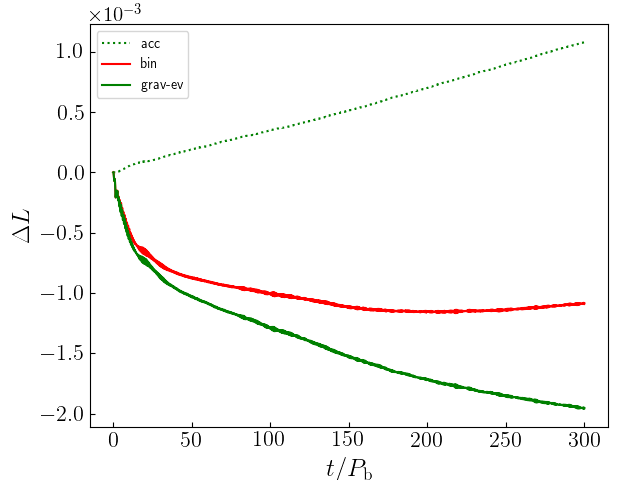}
    \includegraphics[width=0.33\textwidth]{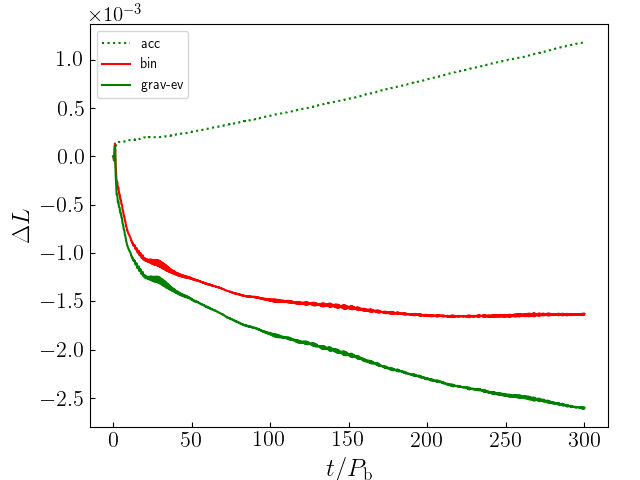}
   \caption{Conservation of angular momentum (see Eq. (\ref{eq:angmomcons2})). The red line shows the binary angular momentum change as calculated from the simulation. The green dotted and solid lines show the accretion and gravity contribution respectively. The left, middle and right panels show the results of the simulations with $H/R=0.05$, $H/R=0.1$ and $H/R=0.2$ respectively. The mass of the disc is $M_{\rm d}=0.1$ for all the runs. 
   }
   \label{fig:momcons}
\end{figure*}

\begin{figure*}
    \centering
    \includegraphics[width=0.4\textwidth]{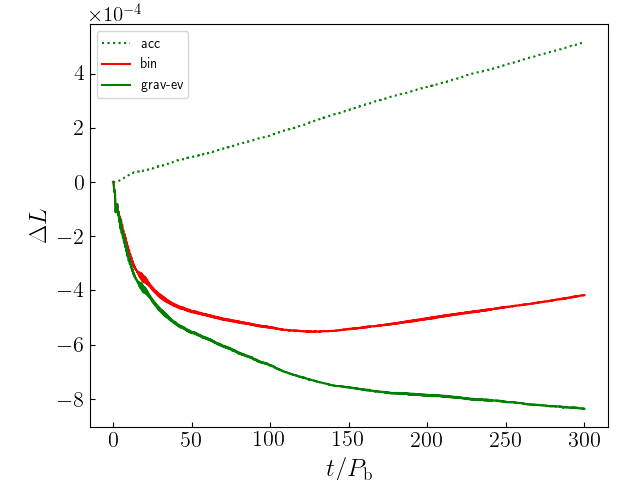}
    \includegraphics[width=0.4\textwidth]{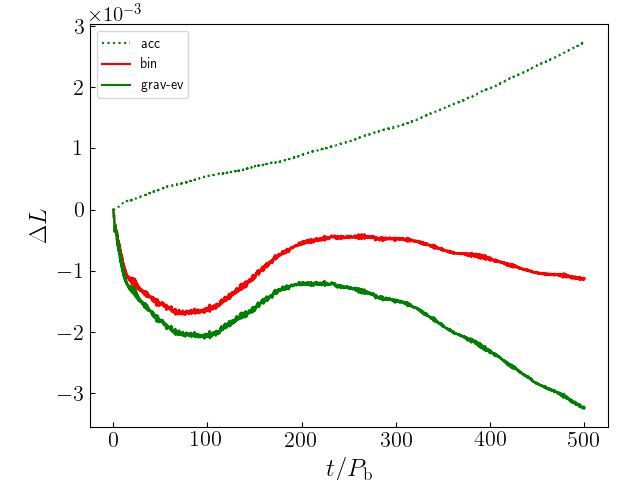}
   \caption{Conservation of angular momentum (see Eq. (\ref{eq:angmomcons2})). The red line shows the binary angular momentum change as calculated from the simulation. The green dotted and solid lines show the accretion and gravity contribution respectively. The left and right panels show the results of the simulations with $M_{\rm d}=0.05$ and $M_{\rm d}=0.2$ respectively. The initial disc aspect ratio is $H/R=0.1$ for both runs.    }
   \label{fig:momcons2}
\end{figure*}

There are two main contributions to the net torque that determine whether the binary shrinks or expands: the gravitational torque exerted by the disc particles onto each individual MBH and the contribution due to accretion of gas particles onto the two MBHs \citep{roedig2012}.

The total angular momentum is exactly conserved in SPH simulations. Therefore we can write

\begin{equation} 
    \frac{d{\bf L}}{dt} = {\bf T}_{\rm G} + \frac{d{\bf L}_{\rm acc}}{dt} 
    \label{eq:angmomcons}
\end{equation}
where ${\bf L}$ is the MBHB angular momentum, the first r.h.s. term is the gravitational torque and the second is the contribution due to accretion onto the MBHs.
Both contributions can be directly computed from the simulations output.

We compute the gravitational torque that the N gas particles exert onto each sink as

\begin{equation}
    {\bf T}_{\rm G} = \sum_{i=1}^{N}{\bf r}_1\times \frac{GM_1m_{\rm i}({\bf r}_{\rm i}-{\bf r}_1)}{|{\bf r}_{\rm i}-{\bf r}_1|^3} + \sum_{i=1}^{N}{\bf r}_2\times \frac{GM_2m_{\rm i}({\bf r}_{\rm i}-{\bf r}_2)}{|{\bf r}_{\rm i}-{\bf r}_2|^3}
    \label{eq:gravt}
\end{equation}
where $m_{\rm i}$ is the particle mass and $r_1,r_2$ are the sinks positions with respect to the binary centre of mass.

The contribution of the accretion of each disc particle to each binary component angular momentum change is instead given by

\begin{equation}
 d{\bf L}_{\rm acc} = {\bf r}_{\rm i} \times m_{\rm i}{\bf v}_{\rm i} - \frac{m_{\rm i}M_{\rm k}}{(m_{\rm i}+M_{\rm k})}\left[({\bf r}_{\rm i}-{\bf r}_{\rm k})\times ({\bf v}_{\rm i}-{\bf v}_{\rm k})\right]
 \label{eq:dLacc}
\end{equation}
where $r_{\rm i},{\bf v}_{\rm i}$ are the position and velocity of the gas particle that is being accreted and $r_{\rm k},{\bf v}_{\rm k}$ are the position and velocity of the sink particle that is accreting. 
The details on the accretion of gas particles onto sinks is outlined in the {\sc phantom} paper \citep{phantom2017}.
The first term on the r.h.s of Eq. (\ref{eq:dLacc}) is the contribution of accretion to the orbital angular momentum of the binary while the second term is the amount of angular momentum that is converted into the spin of the two sink particles.

We can therefore compute the evolution of the binary angular momentum with time directly from the simulations.
The cumulative change due to the gravitational and accretion torque is given by

\begin{equation}
    \Delta {\bf L} = \sum_{dt}\left({\bf T}_{\rm G}dt + d{\bf L}_{\rm acc}\right)
    \label{eq:angmomcons2}
\end{equation}
where we integrate over the whole duration of the simulation. 

Figures \ref{fig:momcons} and \ref{fig:momcons2} show the two contribution to the angular momentum change for each simulation. The dotted lines show the contribution of accretion of gas particles onto the two BHs while the straight lines represent the gravitational torque term in Eq. (\ref{eq:angmomcons2}).
The gravitational torque exerted by the disc onto the binary is always negative and stronger than the torque due to the accretion of material onto the two MBHs in all the simulations we have run.

Whether the circumbinary disc strongly affects the evolution of the binary depends on the location of the resonances and on the disc viscous torque. The former depends on the binary mass ratio and eccentricity while the latter depends on the mechanism that transports angular momentum through the disc, i.e. in our case gravitational instabilities (GI).
The physical origin of these instabilities is related to the standard Jeans instability for homogeneous fluids, where pressure gradients are not efficient in stabilizing large scale disturbances. 
Since the development of these instabilities depends essentially on the ratio between the disc aspect ratio and the disc mass (see Eq. \ref{eq:toomre}) we vary these two quantities in order to explore the parameter space where GIs occur.

\subsection{Effect of the disc aspect ratio}
\label{sec:temp}

\begin{figure}
    \includegraphics[width=\columnwidth]{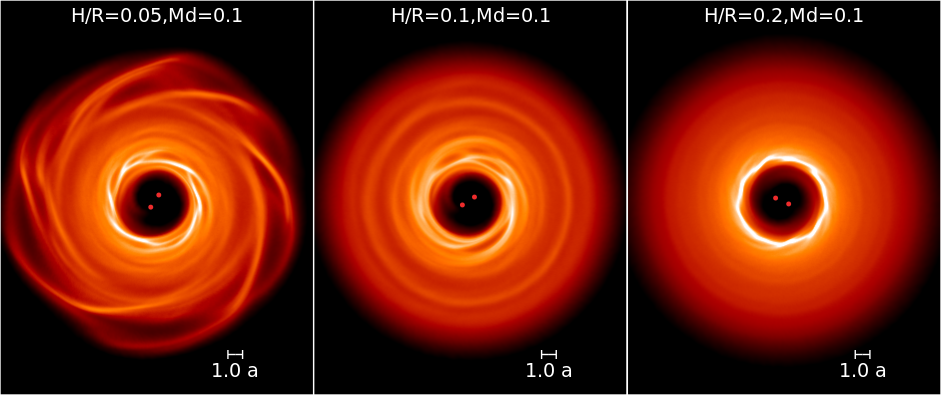}
    \includegraphics[width=\columnwidth]{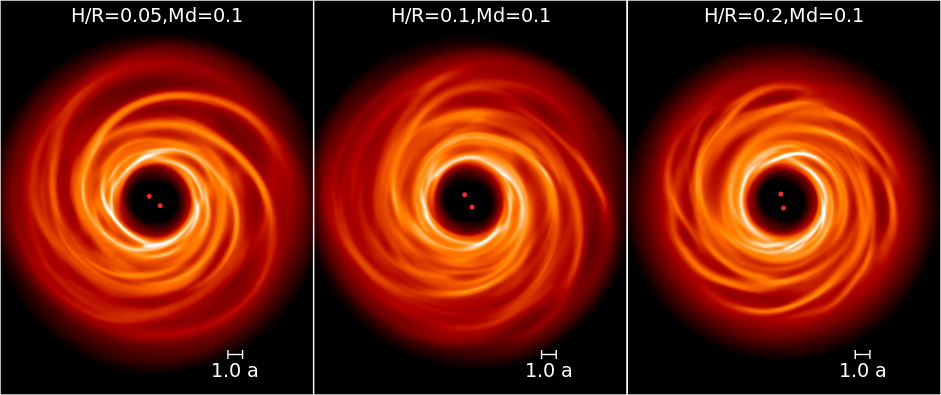}
    \caption{Column density plots of the self-gravitating circumbinary disc around the binary (shown by the red circles).  The view is of the $x$-$y$ plane (i.e. the binary orbital plane) and the density has been integrated through $z$. The color scale spans about two orders of magnitude in density and is the same for all the plots. The color scale spans about two orders of magnitude in density and is the same for all the plots. The three panels correspond to run sg6b, sg3b and sg7b respectively. The upper and lower panels show the result at $t=100,300\,P_{\rm b}$ respectively. }
    \label{fig:temp_splash}
\end{figure}

Previous works have highlighted the importance of the disc temperature in determining the evolution of the binary parameters \citep{tiede2020,heathnixon2020}.
In particular, \cite{heathnixon2020} found that the interaction of non extreme mass ratios binaries with thicker (and therefore warmer) discs, with $H/R\geq 0.2$, results in an increase of the binary semi-major axis because of the large amount of material that is able to enter the cavity with a specific angular momentum larger that that of the binary and to accrete onto the two black holes.

These previous simulations assumed an isothermal equation of state, i.e. the disc temperature remaining constant throughout the simulation.
In order to consider the effects of the disc self-gravity we instead assumed an adiabatic equation of state together with a cooling prescription. 
Self-gravitating accretion disc are expected to self-regulate around a stable value of $\bar{Q}\approx 1$ if the cooling term is not strong enough to drive fragmentation \citep{lodatorice2004,lodato2007sg}.
The Q parameter is proportional to the sound speed, hence increases with the disc temperature, so that cooler discs are more gravitationally unstable.
If a disc has $Q\gg1$ initially and there is no transport or heating mechanism, it eventually cools down due to radiative cooling, until $Q\approx 1$. The disc then develops a GI in the form of a spiral structure. This instability leads to efficient energy dissipation through compression and shocks of the disc material. As a consequence, the disc temperature increases and therefore the value of $Q$ increases again towards a stability value above unity.
The disc is therefore expected, for $\beta_{\rm cool}$ large enough not to be in the fragmentating regime, to be close to marginal stability.
Our simulations correctly reproduce this behaviour.
Figure \ref{fig:temp_splash} shows the snapshots taken after $300\,P_{\rm b}$ of the simulations with different disc initial aspect ratios.
We can see that the structure of the discs is quite different at $100P_{\rm b}$ because of the different initial conditions. However, after $300P_{\rm b}$ the discs have reached very similar spiral patterns and also the cavity size is similar.  

\begin{figure}
    \centering
    \includegraphics[width=0.8\columnwidth]{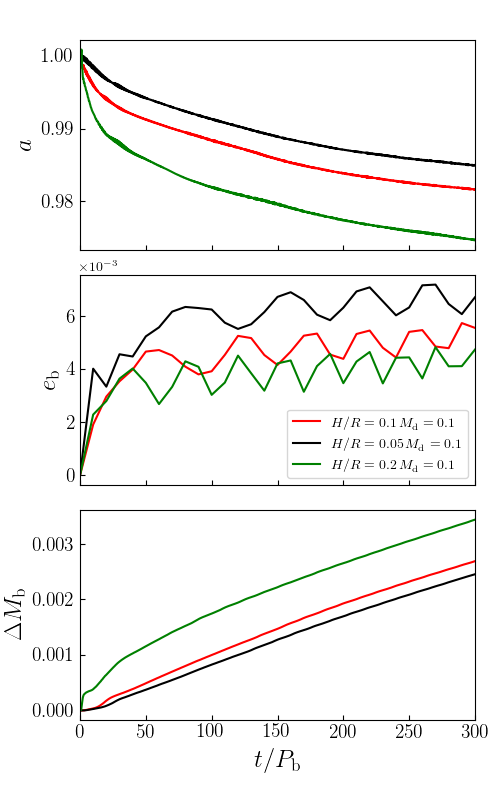}
    \caption{Binary semi-major axis, eccentricity and mass evolution with time, in units of the binary orbital period. The red, black and green lines represent the simulations sg3b, sg6b and sg7b respectively (see Table \ref{tab:sim}).}
    \label{fig:temp_params}
\end{figure}

Figure \ref{fig:temp_params} shows that accretion discs with larger initial aspect ratios drive the binary semi-major axis to decrease on a shorter timescale within the first few tens of orbits. 
The presence of outer Lindblad resonances within the circumbinary disc can lead to the extraction of angular momentum from the binary to the disc \citep{lynden-bell1972} through the excitation of the disc orbits eccentricity, which in turn results in a wave launched outwards. The eccentricity of these orbits can either be damped by viscosity, resulting in circular orbits with increased angular momentum \citep{Lin1979} or by non-linear effects \citep{lubow1998}. 
Since the viscous torque is proportional to $(H/R)^2$ \citep{ss1973}, thicker discs will deliver material at the resonances locations more efficiently compared to thinner discs. Therefore the removal of angular momentum from the binary orbit is faster in warmer discs. 
Note that this behaviour has also been observed by \cite{heathnixon2020} in the case of isothermal non-self-gravitating circumbinary discs with aspect ratios $H/R<0.2$ (see their Fig. 4).
The above is valid in the regime where the disc is not warm enough to drive binary expansion through the capture torque \citep{heathnixon2020}. 

\begin{figure}
    \centering
    \includegraphics[width=0.9\columnwidth]{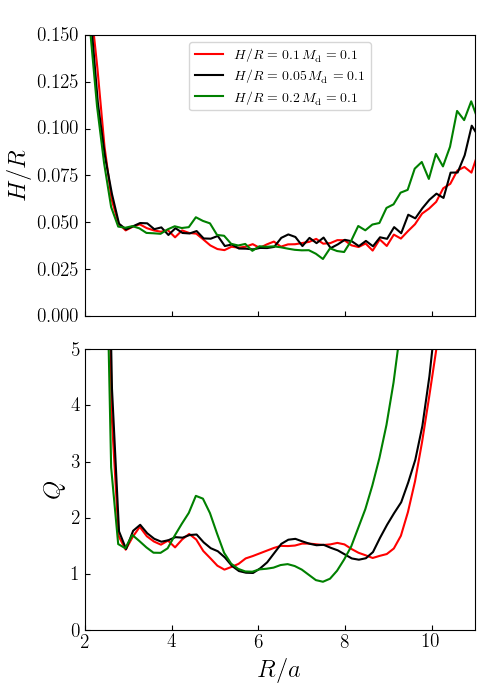}
    \caption{Disc aspect ratio (upper panel) and Q parameter (lower panel) profiles for the circumbinary discs in run sg6b, sg3b and sg7b, represented by the black, red and green line respectively, evaluated at $300P_{\rm b}$. }
    \label{fig:temp_prof}
\end{figure}

Figure \ref{fig:temp_prof} shows that at $300\,P_{\rm b}$, these discs with the same initial mass reach roughly the same temperature (i.e. same aspect ratio), therefore they have very similar Q values. 
This is consistent with Fig. \ref{fig:temp_params} which shows that after an initial transient which is steeper for larger disc aspect ratios, the binary semi-major shrinking rate becomes largely independent from the initial value of the disc temperature.
Note that we observe a decrease in binary semi-major axis even in the simulation that start with $H/R=0.2$ because the disc cools down within the first few orbits and therefore it enters the regime of binary shrinkage \citep{heathnixon2020}.

From the investigation performed in this Section, a general picture emerges in which the long term evolution of the system is independent on the disc initial temperature (i.e. the initial aspect ratio). This is because the dynamical evolution of the disc is determined by the balance between the implemented cooling prescription and heating provided by GIs through the formation of large scale spirals. If the mass (and hence the surface density) of the discs is the same, the properties of the developing spirals would be similar, regardless of the initial aspect ratio, thus providing a comparable source of heating in all three runs, as confirmed by the disc appearance shown in Figure \ref{fig:temp_splash} and by the final $H/R$ and $Q$ parameters shown in Figure \ref{fig:temp_prof}. Crucially, the bulk of the relaxed disc has $0.03<H/R<0.05$, settling onto a regime that promotes binary shrinking.

\subsection{Effect of the initial disc mass}
\label{sec:mass}

\begin{figure}
    \includegraphics[width=\columnwidth]{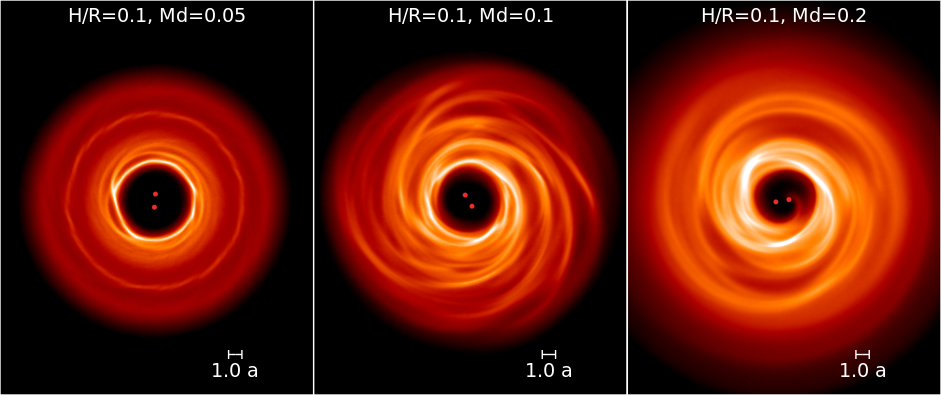}
    \caption{Column density plots for the simulations run sg4b (first panel), sg3b (second panel) and sg8b (third panel) at $t=300\,P_{\rm b}$ respectively for the upper and lower panel.  The view is of the $x$-$y$ plane (i.e. the binary orbital plane) and the density has been integrated through $z$. The color scale spans about two orders of magnitude in density and is the same for all the plots.}
    \label{fig:mass_splash}
\end{figure}

Since the initial disc temperature (i.e. disc aspect ratio) seems to be essentially irrelevant for the evolution of massive discs regulated by self gravity, we now investigate the effect that different initial disc masses have on the evolution of the binary parameters.
Changing the disc mass affects the stability of the disc with respect to GIs. In particular it  changes the value of the disc temperature that sets the self-regulation of the disc, which is expected to change the aspect ratio of the relaxed disc and in turn might have an important effect on the binary shrinking.

\begin{figure}
    \centering
    \includegraphics[width=0.8\columnwidth]{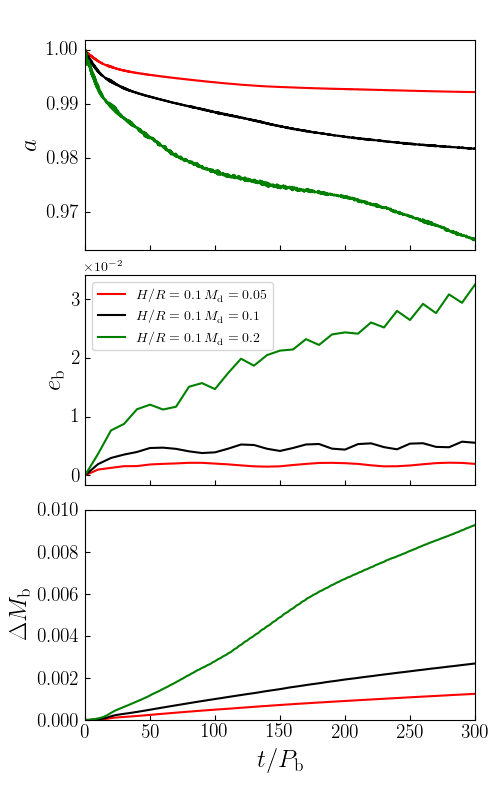}
    \caption{Binary semi-major axis, eccentricity and mass evolution with time, in units of the binary orbital period. The black, red and green lines represent the simulations sg4b, sg3b and sg8b respectively (see Table \ref{tab:sim}).}
    \label{fig:mass_par}
\end{figure}

Figure \ref{fig:mass_splash} shows the snapshots taken after $300\,P_{\rm b}$ of the simulations with different disc initial masses $M_{\rm d}=0.05,\,0.1,\,0.2\,M$ (from right to left panel).
The less massive disc has to cool down further in order to become unstable against GIs. Therefore the transport of angular momentum outwards due to GIs is less efficient. This is consistent with this disc having a larger cavity compared with the other two more massive discs. 
Increasing the initial disc mass leads to a more unstable initial state since we keep the same initial disc temperature.
In the more massive disc case we explored, i.e. rightmost panel in Fig. \ref{fig:mass_splash}, GIs are able to transport angular momentum outwards more efficiently, leading to a smaller cavity and in turn a stronger gravitational interaction with the binary since the material is delivered at the resonances locations on a shorter time scale.
This leads to a much faster initial binary semi-major axis decrease (green line in Fig. \ref{fig:mass_par}) with respect to the less massive discs.

\begin{figure}
    \centering
    \includegraphics[width=\columnwidth]{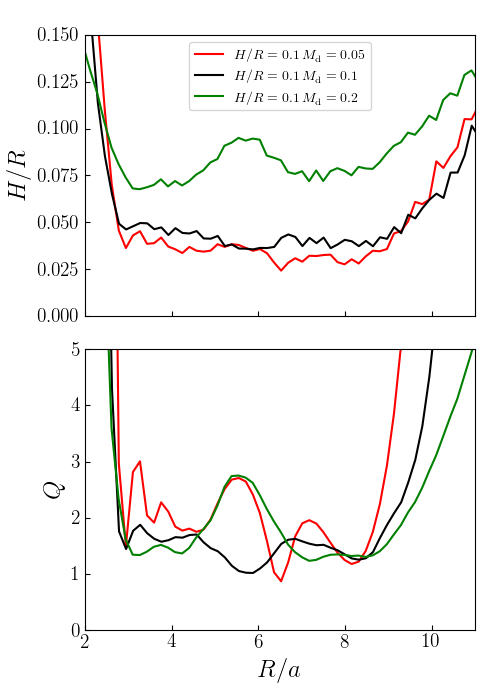}
    \caption{Disc aspect ratio and Q parameter profiles for the circumbinary discs in run sg4b, sg3b and sg8b, represented by the black, red and green line respectively, evaluated at $300P_{\rm b}$. }
    \label{fig:mass_prof}
\end{figure}

Note that in the simulation with initial disc mass $M_{\rm d}=0.2$M, the binary shrinking rate becomes slightly steeper after roughly 200$P_{\rm b}$. This corresponds to the point after which the binary eccentricity increases significantly, reaching a value of $0.03$ at 300$P_{\rm b}$. This is also reflected in the accreted mass onto the binary (bottom panel of Fig. \ref{fig:mass_par}).
The right panel of Fig. \ref{fig:momcons2} shows the same behaviour in terms of the binary angular momentum change as a function of time. 
We found that the binary eccentricity reaches $e=0.05$ while its semi-major axis decreases by about 6\% after 500$P_{\rm b}$ (not shown in the Figures).

From Fig. \ref{fig:mass_prof} we can see that more massive discs reach higher temperatures compared to less massive discs in order to settle onto $Q\approx1$. Nonetheless, $H/R\sim0.07-0.08$ in the bulk of the relaxed disc, which is still small enough to shrink the binary.

\subsection{Effect of the cavity size}
\label{sec:cav}

We also explored the effect that a different initial disc inner edge has on the binary evolution. We changed the inner radius to $R_{\rm in} =3a$ so that this is not too large to slow down the simulations and we do not start with material inside the tidal truncation radius.

\begin{figure}
    \centering
    \includegraphics[width=0.9\columnwidth]{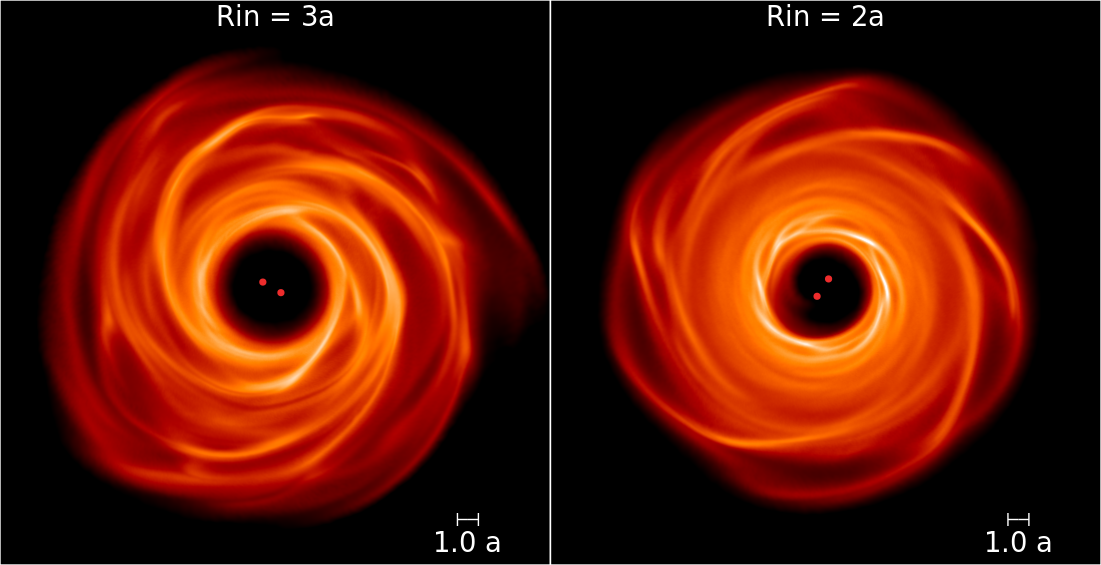}
    \includegraphics[width=0.9\columnwidth]{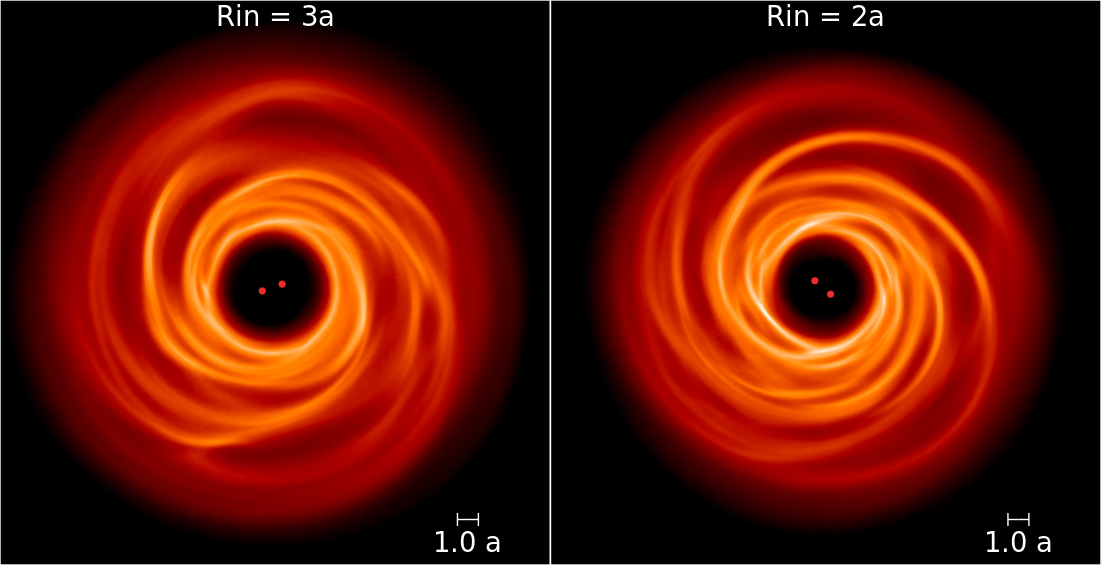}
    \caption{Column density plots for the simulations sg6 (left panels) and sg6b (right panels) at $t=100,\,300\,P_{\rm b}$ respectively for the upper and lower panels.  The view is of the $x$-$y$ plane (i.e. the binary orbital plane) and the density has been integrated through $z$. The color scale spans about two orders of magnitude in density and is the same for all the plots.}
    \label{fig:cav_splash}
\end{figure}

From figure \ref{fig:cav_splash}, we see that the structure of the disc is quite different after $100P_{\rm b}$ and becomes more similar after $300P_{\rm b}$. However, the cavity size remains slightly smaller in the simulation that starts at $2a$ and the spirals are slightly more tightly wrapped as a result of the stronger gravitational interaction with the binary.
The cavity size in the simulation that starts at $3a$ is expected to reach the same value although on a longer timescale because of the transport of angular momentum outwards driven by the spiral arms. 

Note that \cite{heathnixon2020} started their discs at $3a$ and observed a fast viscous spreading inwards because of their choice of $\alpha=0.3$. 
In our case  the angular momentum transport within the disc is regulated by GIs. If the disc is in thermal equilibrium and we assume the heating to be generated by viscous processes only, then the viscosity coefficient and the cooling timescale satisfy \citep{pringle1981,lodato2007sg}

\begin{equation}
    \alpha = \frac{4}{9}\frac{1}{\gamma(\gamma-1)\beta_{\rm cool}}
\end{equation}
which for our choice of cooling time is $\alpha=0.04$.
This is an order of magnitude smaller than the value used by \cite{heathnixon2020}, therefore our disc is expected to viscously evolve on a much longer timescale. In particular, we find that the binary semi-major axis decreases by about 0.1\% in $300$ binary orbits if the disc remains truncated at around $3a$ while the shrinking rate is of about one order of magnitude larger if the disc is truncated at $2a$ (see black line in Fig. \ref{fig:temp_params}).
The disc aspect ratio does reach roughly the same value for a smaller initial cavity, i.e. $H/R\sim0.05$. Therefore, 
for larger initial cavities, the equilibrium is achieved at small enough aspect ratios to lead to binary shrinking, but the evolution is slower.

\section{Discussion and conclusions}
\label{sec:discussion}

The main result of our analysis is that the binary shrinks as a result of the interaction with its self-gravitating circumbinary disc for a variety of initial disc mass and temperature values. 

We have investigated the effect of the initial disc aspect ratio, i.e. disc temperature, on the interaction between the binary and the disc. 
We found that initially thicker discs drive the binary semi-major axis to decrease on a faster timescale. After this initial transient, all the disc with the same initial mass reach the same temperature because of self-regulation and therefore the binary semi-major axis decreases at a rate that is independent on the disc thickness (see Fig. \ref{fig:temp_params}).

Since the stability of a self-gravitating disc against GIs is controlled also by its mass, we evolved our circumbinary discs starting with different initial masses.
More massive discs result in a stronger gravitational interaction with the binary, whose semi-major axis decreases on an initially shorter timescale (see Fig. \ref{fig:mass_par}. 
Interestingly, we see that for the more massive disc case we ran $M_{\rm d}=0.2$M, the binary semi-major axis starts decreasing on a faster timescale after roughly 200$P_{\rm b}$. This corresponds to the point after which the binary eccentricity growth and accreted mass rate do change significantly.
We can therefore conclude that slightly eccentric binaries $e\simeq0.03$ do shrink as a result of the interaction with a relatively thick ($H/R\sim0.08$) circumbinary disc. 
However, the evolution of eccentric binaries deserves a separate investigation and it will be the subject of a future work.

Finally, we also explored the effect of the initial disc inner edge on the binary evolution. We find that discs that start with a smaller cavity size, i.e. close to the truncation radius, drive the binary semi-major axis to shrink on a faster timescale compared to discs that start with larger cavities. Note however that the evolution of the cavity size is driven by viscous processes and therefore, for self-gravitating discs, depends on the choice of the disc parameters that regulate the development of GIs and on the choice of the cooling function.

Note that we cannot resolve the dynamics of the accretion discs that are expected to form around each component of the binary. 
The reason is twofold as, first, our self-gravitating discs regulate themselves at temperatures for which there is little mass able to enter the cavity and therefore we are essentially always in the limit $f\gg1$ \citep{nixonpringle2020}. Secondly because we assumed an adiabatic equation of state within the cavity. \cite{roedig2012} showed that, in similar systems, the use of an isothermal eos inside the cavity does not alter significantly the evolution of the binary semi-major axis with time.
A more robust assessment of the effect of minidiscs on the binary dynamics would require a more physically motivated thermodynamics for the gas in the massive black holes proximity and is therefore deferred to future studies.

Although our main conclusions about the binary evolution are robust, they are valid within the assumptions made to model the disc. Most importantly, we assumed the cooling rate to be proportional to the dynamical timescale without making any assumptions about the physical process that actually drives the disc temperature to decrease. The extension of this work to different types of cooling functions will be the subject of a follow up.
Different cooling prescriptions will essentially change the self-regulation mechanism of self-gravitating accretion discs, which in turn might affect the picture outlined in Section \ref{sec:temp}. A more detailed exploration of the effect of longer cooling timescales and/or different cooling prescriptions will be the subject of a future work.

Finally, we only considered equal mass circular binaries. Since the locations and magnitude of the resonant torques is determined by the mass ratio and eccentricity of the binary it is important to explore different values of these parameters to have a more comprehensive view of the binary-disc interaction. The exploration of these parameters will be the subject of a future work.

\section*{Acknowledgements}

We warmly thank Alessandro Lupi and Chris Nixon for very useful discussions. 
We thank Daniel Price for providing the {\sc phantom} code for SPH simulations and acknowledge the use of {\sc splash} \citep{Price2007} for the rendering of the figures.
AF and AS acknowledge financial support provided under the European Union’s H2020 ERC Consolidator Grant "Binary Massive Black Hole Astrophysics" (B Massive, Grant Agreement: 818691).
We acknowledge the CINECA cluster for providing resources for some of our simulation runs.

\section*{Data availability}

Hydrodynamic simulations used the {\sc phantom} code which
is available from https://github.com/danieljprice/phantom. The input files for generating the
SPH simulations will be shared on reasonable request to the corresponding author.



\bibliographystyle{mnras}
\bibliography{references} 


\bsp	
\label{lastpage}
\end{document}